\documentclass[pra,aps,twocolumn,amsmath,amssymb,superscriptaddress]{revtex4-1}
\usepackage{epsfig,amsmath}
\usepackage{subfigure}
\usepackage{graphicx}
\usepackage{dcolumn}
\usepackage{stmaryrd}
\usepackage{mathrsfs}
\usepackage{pifont}
\usepackage{amsthm}
\usepackage{amssymb}
\usepackage{bm}
\usepackage{latexsym}
\usepackage[colorlinks=true,linkcolor=blue,citecolor=blue]{hyperref}
\usepackage{color}
\usepackage{epstopdf}
\usepackage[ruled]{algorithm2e}

\begin{document}

\title{Optimizing measurement-based cooling by reinforcement learning}

\author{Jia-shun Yan}
\affiliation{School of Physics, Zhejiang University, Hangzhou 310027, Zhejiang, China}

\author{Jun Jing}
\email{Email address: jingjun@zju.edu.cn}
\affiliation{School of Physics, Zhejiang University, Hangzhou 310027, Zhejiang, China}

\date{\today}

\begin{abstract}
Conditional cooling-by-measurement holds a significant advantage over its unconditional (nonselective) counterpart in the average-population-reduction rate. However, it has a clear weakness with respect to the limited success probability of finding the detector in the measured state. In this work, we propose an optimized architecture to cool down a target resonator, which is initialized as a thermal state, using an interpolation of conditional and unconditional measurement strategies. An optimal measurement-interval $\tau_{\rm opt}^u$ for unconditional measurement is analytically derived for the first time, which is inversely proportional to the collective dominant Rabi frequency $\Omega_d$ as a function of the resonator's population in the end of the last round. A cooling algorithm under global optimization by the reinforcement learning results in the maximum value for the cooperative cooling performance, an indicator to measure the comprehensive cooling efficiency for arbitrary cooling-by-measurement architecture. In particular, the average population of the target resonator under only $16$ rounds of measurements can be reduced by four orders in magnitude with a success probability about $30\%$.
\end{abstract}

\maketitle

\section{Introduction}

Cooling mesoscopic and microscopic resonators down to their minimum-energy state is fundamental to observe classical-quantum transition and to exploit quantum advantage in nanoscience~\cite{Nanoscience,OptoReview}. The ground-state preparation is a crucial and implicit step in quantum information processes, including but not limited to the continuous-variable quantum computations~\cite{ContinuousVariables,YouSuperconductingCircuits,BosonSampling,PhononArithmetic}, the ultrahigh precision measurements~\cite{ForceMeasureByOscillator1,ForceMeasureByOscillator2}, and the quantum interface constructions~\cite{CoolingMagnon}. Various strategies are designed to reach an effective temperature as low as possible in the trapped atom and ion systems~\cite{SidebandCooling,SelfCooling,BeatingSidebandCooling}. In atomic laser cooling, popular strategies are consisted of the laser Doppler cooling~\cite{CoolingMagnon,LaserCooling2013,LaserCooling1995}, the resolved-sideband cooling, and the electromagnetically induced transparency (EIT) cooling~\cite{EITCoolingTheory,EITCoolingExperiment}.

Beyond the paradigms extracting the system energy through dissipative channels based on the blue-shifted (anti-Stokes) sidebands, a versatile approach to cooling the mechanical states of motion is provided by the interaction with electromagnetic radiation or quantum measurement. Back-action-evading measurement techniques that can surpass the standard quantum limit have attracted enormous interests. Through pulsed measurement process in optomechanics~\cite{CoolingByPulses,CoolingByPulsesExp,StateSwapByPulses,MotionMeasurementControl,StroboscopicControl}, they can dramatically change the mechanical thermal occupation with no initial cooling. A genuine quantum mechanical cooling engine is proposed~\cite{MeasurementFueledEngine}, whereby the fuel is the energy exchanged with an apparatus performing invasive quantum measurements.

Among these measurement-based techniques, quantum state engineering based on measurements on ancillary systems have been proposed recently in theory~\cite{Purification,OneModeCooling} and demonstrated in experiment~\cite{MeasurementCoolingExp}. Rather than directly detecting the target system, a net nonunitary propagator is realized by inserting projective measurements on the ground state of the detector system in between the joint unitary-evolution segments of target and detector. The induced postselection of the ground state of the target system (typically modelled as a resonator) reduces its high-energy distribution in the ensemble. In another word, the resonator is gradually steered by the outcomes of the conditional measurement (CM) to its ground state via dynamically filtering out its vibrational modes. Ranging from cooling the nonlinear mechanical resonators~\cite{CoolingNonlinearOscillator}, cooling by one shot measurement~\cite{OneShotMeasurement}, expanding cooling range by an external driving~\cite{ExternalLevelCooling}, to accelerating cooling rate by optimized measurement intervals~\cite{TwoModeCooling}, an unexplored weakness of the CM strategies is their limited success probability inherited from the projective operation. An amount of experimental overhead rises unavoidably with more samples in ensemble. In sharp contrast to CM, the unconditional measurement (UM) strategy performs a nonselective and impulsive measurement in all the bases of the bare Hamiltonian of detector in the end of each round of the joint evolution~\cite{CoolingQubit,FockState}. It is more likely to realize a unit-success-probability cooling but suffers from a much slower cooling rate than CM, indicating much more number of measurements toward the ground-state cooling. To compromise the cooling rate and the success probability, the interpolating-configuration of conditional and unconditional measurements constitutes an optimization problem.

The integration of a small-scale quantum circuit with a classical optimizer, e.g., the neural network, provides a paradigm by designing a sequence of parametrized quantum operations that are well suited to implement robust and high-fidelity algorithms. Many reinforcement learning (RL) algorithms constructed by the neural network, that demonstrated remarkable capabilities in the board and video games~\cite{GameGoNN,GameGoWithoutHuman,GameChess,HumanLevelControl}, have substantiated a widely and timely interest in studying quantum physics~\cite{MLAndPhysics}, such as quantum error correction~\cite{ContinuousErrorCorrection,ErrorCorrectionFeedback}, quantum simulation~\cite{DigitalQuantumSimulation,SimulationHybridNetwork}, and quantum state preparation~\cite{PhaseTransition,DifferentPhaseControl,StatePreparationRL}, to name a few. The proximal policy optimization (PPO) algorithm, as a typical RL algorithm with a significant sample complexity, scalability, and robustness for hyperparameters, has proven to be a fruitful tool in quantum optimization control~\cite{ModelFreeControlPPO,FeedbackControlPPO,ManyBodyStatePreparationPPO}.

In this work, we propose a measurement-based cooling architecture as a hybrid sequence of UM and CM strategies. It involves a double optimization: for each step along the sequence, either UM or CM can be considerably improved by using a local optimized measurement interval; and for the global efficiency of the sequence, its arrangement can be separably optimized through reinforcement learning. Particularly, in a typical measurement-based cooling model, i.e., the Jaynes-Cummings (JC) model, where a mechanical resonator (the target system) is coupled to a qubit (the detector system), conditional and unconditional measurements are alternatively performed to cool down the resonator to its ground state. A feedback scheme is triggered upon calling a CM to determine whether or not to launch the next round of evolution-and-measurement according to the measurement outcome. Analogous to the optimized measurement-interval obtained for CM~\cite{TwoModeCooling}, we analytically derive an optimized interval for UM. Then the free-evolution intervals between any neighboring measurements, either UM or CM, can be optimized for cooling. The global sequence of measurements or the implementing order of UM and CM can be further optimized with reinforcement learning. The optimizer is fed with the cooperative cooling performance, a function of the average population of the resonator, the success probability of the detector in the measured subspace, and the fidelity of the resonator in the ground state. Eventually we find an optimal sequence holding an overwhelming advantage over all the others.

The rest of this work is structured as follows. We briefly revisit the general framework for the cooling protocols based on conditional and unconditional measurements in Secs.~\ref{CM} and \ref{UM}, respectively. In Sec.~\ref{UM}, an analytical expression of the optimized measurement-interval is obtained for UM. In Sec.~\ref{Optimization}, we introduce the interpolation diagram for the cooling architecture based on these two measurements, define the cooperative cooling performance to comprehensively quantify various strategies, and present the optimized result through reinforcement learning. The PPO algorithm and the optimal-control procedure are provided in Appendixes~\ref{PPOSec} and \ref{OptSequence}, respectively. The whole work is discussed and summarized in Sec.~\ref{Conclusion}.

\section{Conditional and unconditional measurements}\label{CMandUM}

\subsection{Conditional Measurement}\label{CM}

Consider a JC model used for cooling-by-measurement protocols, whose Hamiltonian in the rotating frame with respect to $H_0=\omega_a(|e\rangle\langle e|+a^\dagger a)$ reads
\begin{equation}\label{Ham}
H=\Delta|e\rangle\langle e|+g(a^\dagger\sigma_-+a\sigma_+).
\end{equation}
Here $\Delta\equiv\omega_e-\omega_a$ is the detuning between the level-spacing of the atomic detector $\omega_e$ and the frequency of the target resonator $\omega_a$ and $|\Delta|\ll\omega_e, \omega_a$. $g$ is the coupling strength between the detector (qubit) and the target resonator. Pauli matrices $\sigma_-$ and $\sigma_+$ denote the transition operators of the qubit; and $a$ ($a^\dagger$) represents annihilation (creation) operator of the resonator.

The conditional measurement-based cooling is described by a sequence of piecewise joint evolutions of the resonator and the detector, that are interrupted by instantaneous projective measurements on a particular subspace of the detector. Initially, the resonator is in a thermal-equilibrium state $\rho_a^{\rm th}$ with a finite temperature $T$, and the detector qubit starts from the ground state. Then the overall initial state has the form of $\rho_{\rm tot}(0)=|g\rangle\langle g|\otimes\rho_a^{\rm th}$. To cool down the resonator, a conditional or selective measurement $M_g=|g\rangle\langle g|$ is implemented on the detector after the free-evolution with an interval $\tau$, when the overall state becomes $\rho_{\rm tot}(\tau)=\exp(-iH\tau)\rho_{\rm tot}(0)\exp(iH\tau)$. And then conditional measurement yields a probabilistic result:
\begin{equation}
\rho_a(\tau)=\frac{\langle g|\rho_{\rm tot}(\tau)|g\rangle}{{\rm Tr}\left[\langle g|\rho_{\rm tot}(\tau)|g\rangle\right]}.
\end{equation}
Based on the time-dependence of the interval $\tau$, conditional cooling protocols can be categorized into the equal-time-spacing and unequal-time-spacing strategies~\cite{OneModeCooling,TwoModeCooling}. The unequal-time-spacing strategy has demonstrated a dramatic cooling efficiency by setting the measurement interval as the inverse of the time-evolved thermal Rabi frequency $\tau_{\rm opt}^c(t)=1/\Omega_{\rm th}(t)$, where $\Omega_{\rm th}(t)\equiv g\sqrt{\bar{n}(t)}=g\sqrt{\sum_nnp_n(t)}$ with $p_n(t)$ denoting the current population of the resonator on the Fock state $|n\rangle$. To optimize the cooling performance, our cooling architecture in this work employs the unequal-time-spacing strategy. After $N$ rounds of free-evolution and instantaneous-measurement described by an ordered time sequence $\{\tau_1(t_1), \tau_2(t_2), \cdots, \tau_N(t_N)\}$ with $t_{i>1}=\sum_{j=1}^{j=i-1}\tau_j$ and $\tau_1\equiv1/[g\sqrt{{\rm Tr}(\hat{n}\rho_a^{\rm th})}]$, the resonator state becomes
\begin{equation}\label{rhoace}
\rho_a\left(t=\sum_{i=1}^N\tau_i\right)=\frac{\sum_{n}\prod_{i=1}^N|\alpha_n(\tau_i)|^2p_n|n\rangle\langle n|}{P_g(N)},
\end{equation}
where
\begin{equation}
p_n=\frac{e^{-n\hbar\omega_a/k_BT}}{Z}, \quad Z\equiv\frac{1}{1-e^{-\hbar\omega_a/k_BT}}
\end{equation}
is the initial population,
\begin{equation}
P_g(N)=\sum_n\prod_{i=1}^N|\alpha_n(\tau_i)|^2p_n
\end{equation}
is the survival or success probability of CM, and
\begin{equation}\label{CMCoolingCoeff}
\left|\alpha_n(\tau_i)\right|^2=\frac{\Omega_n^2-g^2n\sin^2(\Omega_n\tau_i)}{\Omega_n^2}
\end{equation}
is the cooling coefficient with $\Omega_n=\sqrt{g^2n+\Delta^2/4}$ denoting the $n$-photon Rabi frequency. The cooling coefficient in Eq.~(\ref{rhoace}) determines the average population
\begin{equation}
\bar{n}(t)={\rm Tr}\left[\hat{n}\rho_a(t)\right], \quad \hat{n}\equiv a^{\dagger}a,
\end{equation}
by reshaping the population distributions over all the Fock states. Note in Eq.~(\ref{CMCoolingCoeff}), the cooling coefficient for $|0\rangle$ is unit, $|\alpha_0(\tau_i)|^2=1$, meaning that the ground-state population is always under protection during the cooling process. The populations on high-occupied Fock states are gradually reduced by $|\alpha_n(\tau_i)|^N<1$ with increasing $N$ unless $\sin(\Omega_n\tau_i)=0$ or $\Omega_n\tau_i=j\pi$ with integer $j$.

\subsection{Unconditional Measurement}\label{UM}

Unconditional-measurement cooling is a statistical mixture of the conditional-measurement counterpart, by expanding the measurement subspace to the whole space of the detector system. After a period of joint unitary evolution under the Hamiltonian~(\ref{Ham}), the overall state can be written as
\begin{equation}
\rho_{\rm tot}(\tau)=\bigoplus_np_n\begin{pmatrix}|\alpha_n(\tau)|^2 & \chi_n(\tau)\\
\chi^*_n(\tau)& |\beta_n(\tau)|^2\end{pmatrix},
\end{equation}
where
\begin{equation*}
\begin{aligned}
&\chi_n(\tau)\equiv\frac{-g\sqrt{n}\left[\Delta\sin^2(\Omega_n\tau)-i\Omega_n\sin(2\Omega_n\tau)\right]}{2\Omega_n^2},\\
&|\beta_n(\tau)|^2\equiv\frac{g^2n\sin^2(\Omega_n\tau)}{\Omega_n^2}.
\end{aligned}
\end{equation*}
UM can be implemented by tracing out the degrees of freedom of the detector ${\rm Tr}_d[\rho_{\rm tot}(\tau)]$. Then the resonator state reads
\begin{equation}\label{umrhoa}
\rho_a(\tau)=\sum_{n\geq0}\left[|\alpha_n(\tau)|^2p_n+|\beta_{n+1}(\tau)|^2p_{n+1}\right]|n\rangle\langle n|.
\end{equation}
So that after a nonselective measurement, i.e., a measurement without recording the result, a population transfer in the target resonator occurs as
\begin{equation}\label{pn}
p_n\rightarrow|\alpha_n(\tau)|^2p_n+|\beta_{n+1}(\tau)|^2p_{n+1}.
\end{equation}
In contrast to CM strategy that is characterized by a single cooling coefficient $|\alpha_n|^2$ in Eq.~(\ref{CMCoolingCoeff}), UM strategy depends subtly on an extra cooling coefficient $|\beta_n|^2$. According to Eq.~(\ref{pn}), the initial population on the ground state $p_0$ becomes $|\alpha_0(\tau)|^2p_0+|\beta_1(\tau)|^2p_1=p_0+|\beta_1(\tau)|^2p_1$, indicating that a part of population on the first excited state is transferred to the ground state. Under rounds of nonselective measurements, it is intuitive to expect that the populations on the higher states of the resonator keep moving to the lower states and eventually to the ground state. In practice, the cooling is however constrained and even invertible since the populations on certain excited states can be fixed or enhanced when $|\alpha_n(\tau)|^2=1$ and $|\beta_{n+1}(\tau)|^2\geq0$, i.e., $\Omega_n\tau=1$ and $\Omega_{n+1}\tau\geq0$. This problem can be addressed by employing the unequal-time-spacing strategy. A time-varying $\tau$ could ensure that populations on all excited states are gradually reduced.

\begin{figure}[htbp]
\centering
\includegraphics[width=0.95\linewidth]{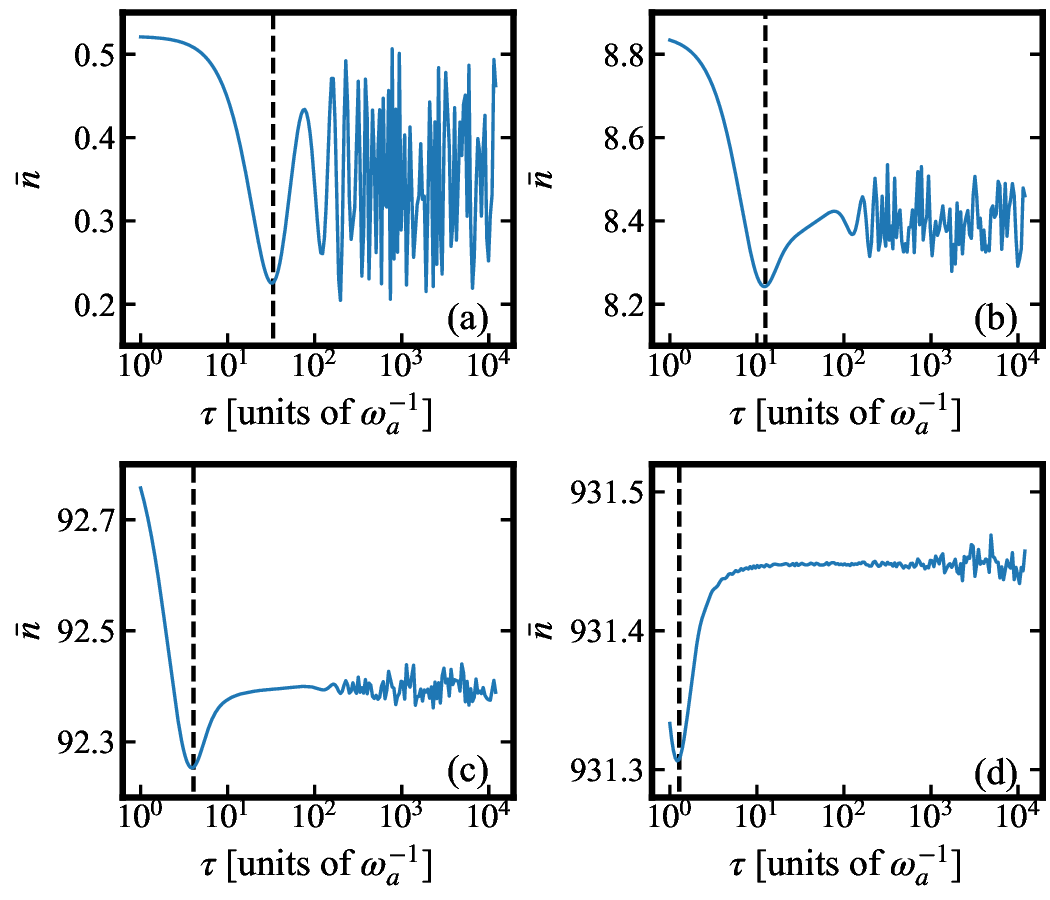}
\caption{Average population of the resonator after a single unconditional measurement as a function of the measurement-interval $\tau$ under various initial temperatures. (a) $T=0.01$ K, (b) $T=0.1$ K, (c) $T=1.0$ K and (d) $T=10$ K. The vertical black-dashed lines indicate the analytical results for the optimized intervals given by Eq.~(\ref{Optimaltau}). The parameters for the blue-solid curves are set as $g=0.04\omega_a$ and $\Delta=0.01\omega_a$. }\label{OptimalInterval}
\end{figure}

Cooling efficiency of UM strategy depends severely on the choice of $\tau$ spacing neighboring measurements, analogous to that of CM~\cite{TwoModeCooling}. That could be observed in Fig.~\ref{OptimalInterval} by the average population of the resonator $\bar{n}$ under one measurement on the detector. The $\tau$-dependence of $\bar{n}$ demonstrates similar patterns across four orders in scale of initial temperature. It is found that the average population declines gradually to a minimal point (the relative reduction becomes smaller as increasing temperature) at an optimized measurement-interval $\tau_{\rm opt}^u$, then rebounds quickly and ends up with a random fluctuation around a value slightly lower than its initial thermal occupation $\bar{n}_{\rm th}\equiv{\rm Tr}(\hat{n}\rho_a^{\rm th})$.

To make full use of the cooling strategy, it is desired to analytically find the optimized interval $\tau_{\rm opt}^u$ as a functional of the current state and the model parameters. By virtue of Eq.~(\ref{umrhoa}) and under the resonant condition, the average population after a single unconditional measurement reads
\begin{equation}\label{nbar}
\begin{aligned}
\bar{n}&=\sum_{n\geq0}n\left(p_n\cos^2\Omega_n\tau+p_{n+1}\sin^2\Omega_{n+1}\tau\right)\\
&=\eta+\frac{1}{2Z}\sum_{n\geq0}ne^{-nx}(\cos2\Omega_n\tau-e^{-x}\cos2\Omega_{n+1}\tau),
\end{aligned}
\end{equation}
where $\eta\equiv(\bar{n}_{\rm th}+2\bar{n}^2_{\rm th})/(2+2\bar{n}_{\rm th})$ and $x\equiv\hbar\omega_a/k_BT$. Since the weight function $ne^{-nx}$ in Eq.~(\ref{nbar}) is dominant around $n_d\equiv k_BT/\hbar\omega_a=1/x$, the variables $\Omega_n$ and $\Omega_{n+1}$ could thus be expanded around $n=n_d$. To the first order of $n-n_d$, we have
\begin{equation*}
\begin{aligned}
&\cos2\Omega_n\tau-e^{-x}\cos2\Omega_{n+1}\tau \\
&\approx\cos2\Omega_{d}\tau-e^{-x}\cos2\Omega_{d+1}\tau+(n-n_d) \\
&\times\left(-\frac{\Omega_d\tau\sin2\Omega_d\tau}{n_d}+e^{-x}\frac{\Omega_{d+1}\tau\sin2\Omega_{d+1}\tau}{n_d+1}\right),
\end{aligned}
\end{equation*}
where
\begin{equation}\label{Omegad}
\Omega_d\equiv g\sqrt{n_d}, \quad \Omega_{d+1}\equiv g\sqrt{n_d+1}
\end{equation}
define the dominant Rabi frequencies. Under the approximations that appropriate for a moderate temperature $e^{-x}=\bar{n}_{\rm th}/(\bar{n}_{\rm th}+1)\approx1$ and $\Omega_{d+1}/(n_d+1)\approx\Omega_d/n_d$, the average population in Eq.~(\ref{nbar}) can be expressed by
\begin{equation}\label{approxn1}
\bar{n}\approx\eta+\sin\Omega_-\tau\left(\bar{n}_{\rm th}\sin\Omega_+\tau+\eta'\Omega_d\tau\cos\Omega_+\tau\right),
\end{equation}
where $\Omega_{\pm}\equiv\Omega_{d+1}\pm\Omega_d$ and $\eta'\equiv\bar{n}_{\rm th}(1+2\bar{n}_{\rm th}-n_d)/n_d$. Note we have used the formulas about the geometric series $\sum_{n=0}^\infty ne^{-nx}=e^x/(e^x-1)^2$ and $\sum_{n=0}^\infty n^2e^{-nx}=e^x(1+e^x)/(e^x-1)^3$. Within a moderate time step $\tau$, Eq.~(\ref{approxn1}) depends predominantly on the high-frequency terms characterized by $\Omega_+$. In the regime of $T\sim 0.1-10$ K, the term weighted by $\eta'\Omega_d\tau$ overwhelms that weighted by $\bar{n}_{\rm th}$. And as evidenced by Fig.~\ref{OptimalInterval}, this advantage expands with a larger $\tau_{\rm opt}^u$ given the initial or effective temperature of the resonator becomes lower. We can therefore focus on the last term in Eq.~(\ref{approxn1}) to minimize $\bar{n}$. Subsequently, $\cos\Omega_+\tau=-1$ yields
\begin{equation}\label{Optimaltau}
\tau_{\rm opt}^u=\frac{\pi}{\Omega_d+\Omega_{d+1}}.
\end{equation}
This result can be extended to the near-resonant situation by modifying the definition of $\Omega_d$ in Eq.~(\ref{Omegad}) to $\sqrt{g^2n_d+\Delta^2/4}$. The vertical black-dashed lines in Fig.~\ref{OptimalInterval} denote the measurement-intervals optimized by Eq.~(\ref{Optimaltau}). It is found that the analytical expression is well suited to estimate the minimum values of average population in a wide range of temperature. As demonstrated by both analytical and numerical results, a shorter measurement-interval is demanded to cool down a higher-temperature resonator. In the JC-like models, coupling a qubit to a high-temperature resonator induces a faster transition between the ground state and the excited state of the qubit. Although a quick measurement would interrupt this process, an unappropriate time-interval would have a negative effect on cooling~\cite{CoolingQubit}. 

Similar to the optimized interval $\tau_{\rm opt}^c(t)$ for the conditional-measurement strategy~\cite{TwoModeCooling}, here $\tau_{\rm opt}^u$ is also updatable by substituting time-varied $\Omega_d$ and $\Omega_{d+1}$ to Eq.~(\ref{Optimaltau}). The dominant Fock-state-number $n_d$ determining $\Omega_d$ in Eq.~(\ref{Omegad}) could be understood as a function of the effective temperature during the cooling procedure, which relies uniquely on $\bar{n}(t)$ or $p_n(t)$.

\section{Measurement optimization}\label{Optimization}

Thermal resonator could be steadily yet slowly cooled down by unconditional measurement strategy equipped with an optimized measurement-interval in Eq.~(\ref{Optimaltau}). And this strategy is performed with a unit probability in the absence of postselection over the measurement outcome. In sharp contrast, conditional measurement strategy is a more efficient cooling protocol but with a poor success probability. It is therefore desired to find an optimized sequence of measurements as a hybrid of UM and CM to hold a great performance taking both cooling efficiency and experimental overhead into account. In this section, we present an algorithm that employs the reinforcement learning to generate the optimized control sequence indicating when and which measurement is performed.

The performance of any cooling-by-measurement strategy can be characterized or evaluated by the cooling ratio $\bar{n}(t)/\bar{n}_{\rm th}$, the success probability $P_g$ of the detector in the measured subspace, and the fidelity of the resonator in its ground state $F=\langle n=0|\rho_a(t)|n=0\rangle$~\cite{OneModeCooling}. To compare various interpolation sequences of UM and CM in cooling performance and to evaluate the figure of merit for the reinforcement learning, we can define a cooperative cooling quantifier as
\begin{equation}\label{cooperative}
\mathcal{C}=FP_g\log_{10}{\frac{\bar{n}_{\rm th}}{\bar{n}(t)}}.
\end{equation}
Notably, the logarithm function is used to obtain a positive value with almost the same order as $F$ and $P_g$ in magnitude. Then $\bar{n}(t)$, $P_g$, and $F$ could be considered in a balanced manner. In fact, the average population could be reduced by several (normally less than $10$) orders in magnitude under an efficient cooling protocol. In the EIT cooling~\cite{EITCoolingExperiment2020}, $\log_{10}[\bar{n}_{\rm th}/\bar{n}(t)]\sim(2, 3)$; and in the resolved sideband cooling ~\cite{SidebandCooling2016}, $\log_{10}[\bar{n}_{\rm th}/\bar{n}(t)]\sim(4, 5)$. Although Eq.~(\ref{cooperative}) is not a unique choice, it is instructive to find that a lower average population, a larger success probability, and a higher ground-state fidelity to yield a better cooling performance.

\begin{figure}[htbp]
\centering
\includegraphics[width=0.95\linewidth]{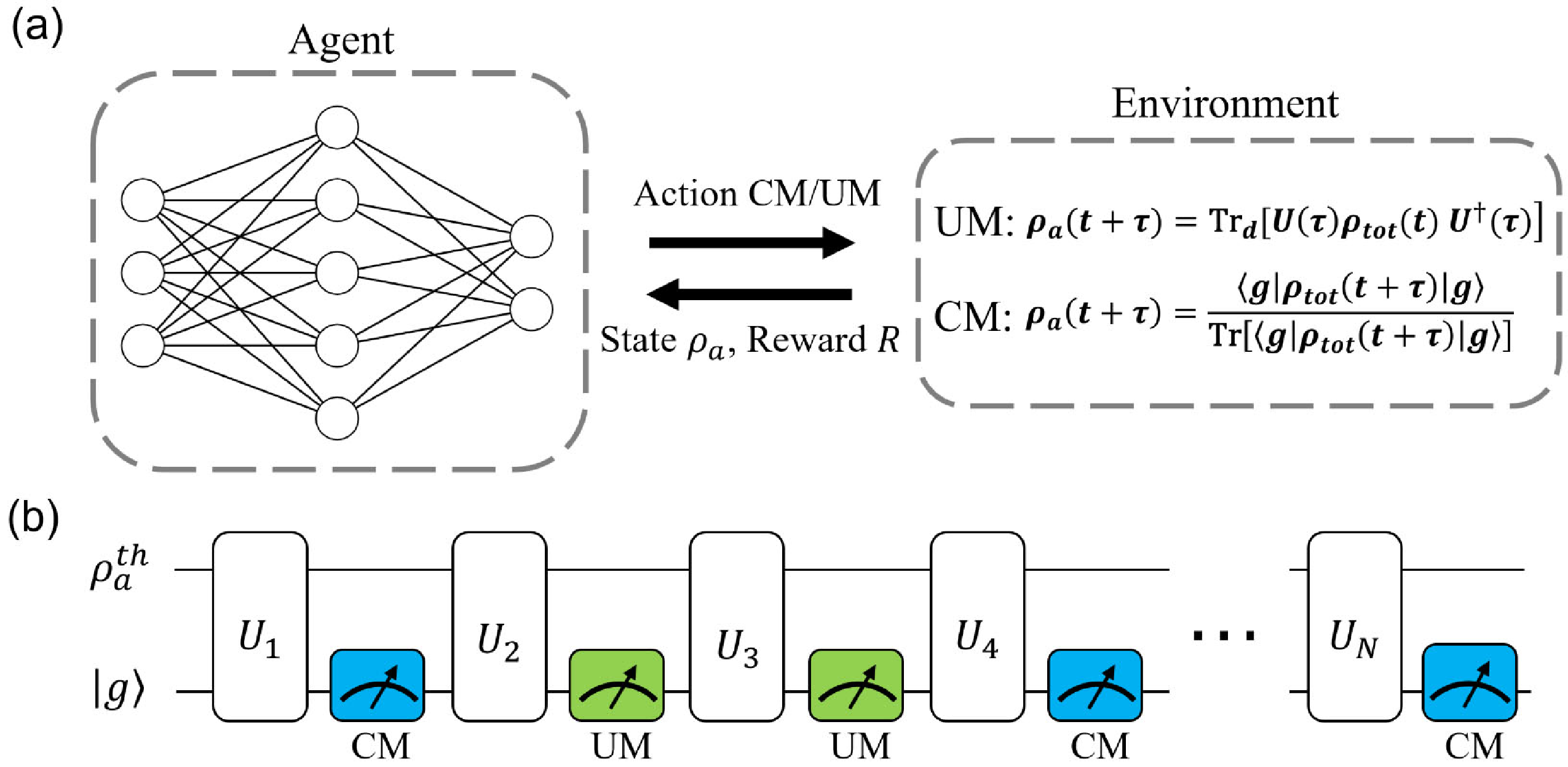}
\caption{(a) RL-optimization diagram on cooling by measurement. An agent constructed by the neural network interacts with an environment. The agent chooses an action (CM or UM strategy) according to the current state of the resonator. Then the environment would take this action and return both the state under the measurement and the reward $R$ based on the cooperative cooling performance $\mathcal{C}$ in Eq.~(\ref{cooperative}). (b) Circuit model for our cooling algorithm based on the local-optimized UM and CM strategies. Starting from a thermal state, the resonator (the upper line) would be gradually cooled down to its ground state with implementation of measurement on detector (the lower line), which starts from the ground state. The measurement sequence can be obtained by the reinforcement learning. }\label{Model}
\end{figure}

The RL-optimization is shown in Fig.~\ref{Model}(a). It is constituted by the ``agent'' part based on a series of neural network and the ``environment'' part performing the cooling-by-measurement actions on quantum system. In the reinforcement learning, the agent has a cluster of parameters, which would be learned and trained using the data collected through its interaction with the environment. In our architecture, the agent would choose an action, i.e., conditional or unconditional measurement, on the resonator, given its current state. Then the environment takes this action and returns the updated resonator-state $\rho_a$ and a ``reward'' $R$ after the measurement. The reward is generated by the indicator in Eq.~(\ref{cooperative}) to estimate whether the action is good or bad, that would be used to update the agent's parameters. During one ``episode'', the agent would interact with the environment for $N$ times, i.e., the number of measurements during the whole sequence, which has been fixed from the beginning. A total reward is eventually counted. And the agent is trained to maximize the total reward through artificial episodes until it converges. Then the agent could provide a realistic control sequence of the measurement strategies with their own (optimized) measurement intervals. The cooling-by-measurement sequence can be realized in a circuit model in Fig.~\ref{Model}(b). Rounds of free-evolutions and measurements are successively arranged. The evolution time between two neighboring measurements depends on the measurement strategy and the resonator state at the end of the last round. We follow the PPO algorithm in the agent structure, the data-collecting methods, and the updating parameters, whose details can be found in Appendix~\ref{PPOSec}. The interpolation algorithm of UM and CM and the implementation of the measurement sequence are illustrated by a pseudocode in Appendix~\ref{OptSequence}.

\begin{figure}[htbp]
\centering
\includegraphics[width=0.95\linewidth]{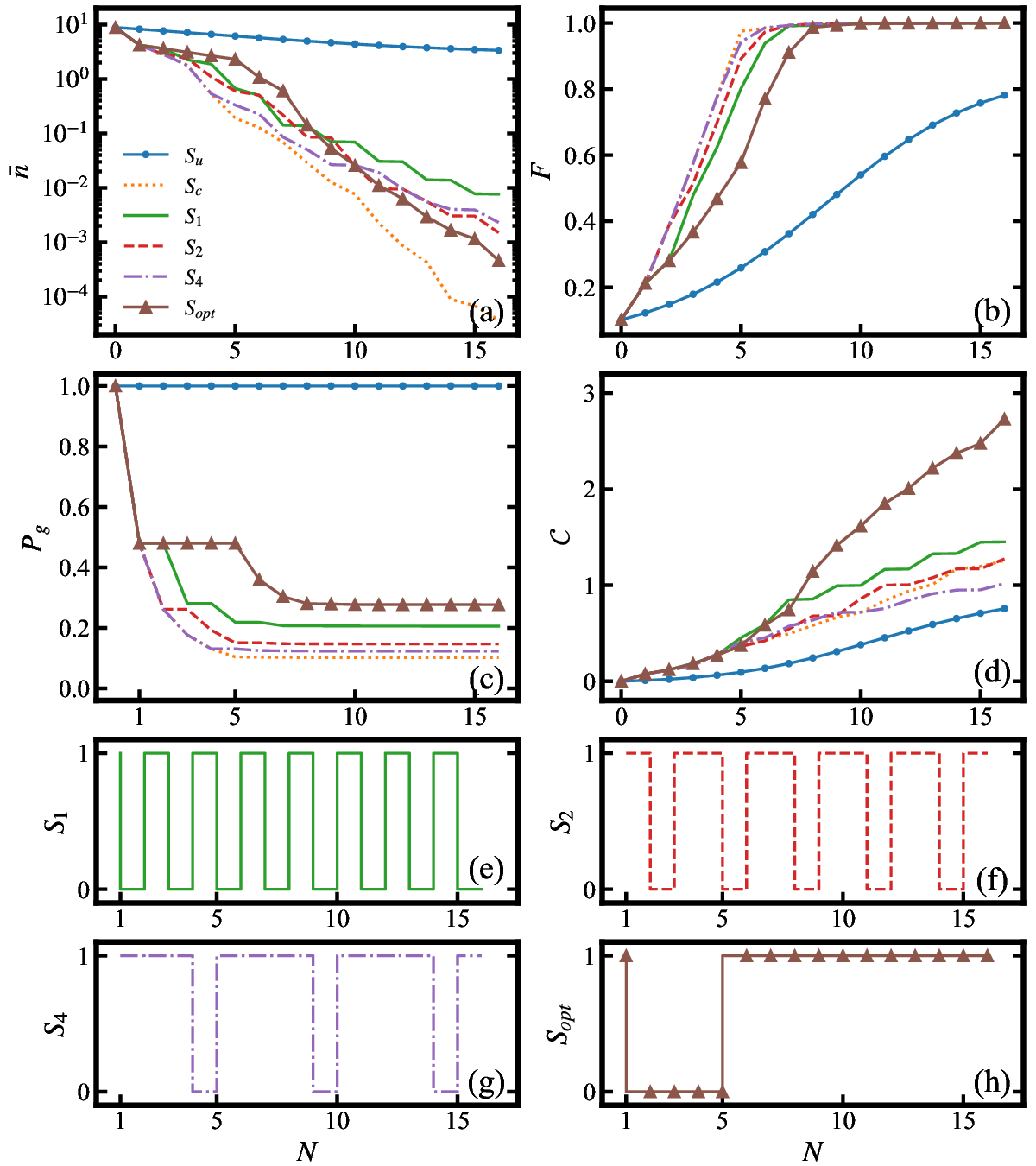}
\caption{(a) Average population, (b) Fidelity of the resonator in its ground state, (c) Success probability, and (d) Cooperative cooling performance under various sequences of cooling-by-measurement. The blue-solid lines with circle markers labeled by $S_u$ and the orange-dotted lines labeled by $S_c$ indicate the sequences entirely consisting of UM and CM strategies, respectively. The green-solid lines, the red-dashed lines, and the purple-dot-dashed lines describe the hybrid sequences shown in (e), (f), and (g), and labeled by $S_1$, $S_2$, and $S_4$, respectively. The brown-solid lines with triangle markers labeled by $S_{\rm opt}$ is the RL-optimized sequence presented in (h). For all the sequences in (e), (f), (g), and (h), $1$ and $0$ indicate CM and UM strategies, respectively. The parameters are set as $\omega_a=1.4$ GHz, $T=0.1$ K, $g=0.04\omega_a$, and $\Delta=0.01\omega_a$. }\label{CoolingPerformance}
\end{figure}

We consider to cool down a mechanical microresonator in gigahertz~\cite{MechanicalResonator1,MechanicalResonator2} with various interpolation sequences of UM and CM. Using the resonator-frequency $\omega_a=1.4$ GHz, the coupling strength between resonator and detector $g=0.04\omega_a$ and the initial temperature of resonator $T=0.1$ K, it is found that the average population starts from $\bar{n}_{\rm th}=8.85$. The cooling performances under the sequences entirely consisting of UM and CM are shown by the blue-solid lines with circle markers and the orange-dotted lines in Figs.~\ref{CoolingPerformance}(a)-(d), labeled by $S_u$ and $S_c$, respectively. It is found that under the conditional measurement strategy with $N=16$, the average population $\bar{n}$ is reduced by five orders in magnitude [see Fig.~\ref{CoolingPerformance}(a)] and the ground-state fidelity is over $F>0.9999$ [see Fig.~\ref{CoolingPerformance}(b)] with less than $10\%$ of the success probability [see Fig.~\ref{CoolingPerformance}(c)]. In sharp contrast, under the same number of unconditional measurements, $\bar{n}$ is merely reduced to $\bar{n}\approx 3.36$ and the ground-state fidelity $F\approx0.78$, despite with a unit success probability. In terms of all the individual quantifiers, i.e., $\bar{n}$, $F$, and $P_g$, the results under the hybrid sequences of UM and CM labelled by $S_k$, $k=1,2,4$, are among the former two limits $S_u$ and $S_c$. As illustrated by Figs.~\ref{CoolingPerformance}(e), (f), and (g), the three sequences start from a CM (indicated by $1$), switch to the UM (indicated by $0$) after $k$ rounds of free-evolution and measurement, switch back to CM after a single round, and then repeat the preceding arrangement. In comparison to the entire UM sequence, the interpolation with CM promotes the cooling efficiency in $\bar{n}$. A larger $k$ gives rise to a smaller proportion of the unconditional measurements and a less probability $P_g$ that the detector remains in its measured subspace.

With respect to the cooperative cooling performance given by Eq.~(\ref{cooperative}), it is found [see Fig.~\ref{CoolingPerformance}(d)] that $\mathcal{C}(S_1)>\mathcal{C}(S_2)>\mathcal{C}(S_4)>\mathcal{C}(S_u)$ and yet $\mathcal{C}(S_2)\approx\mathcal{C}(S_c)$. Such that a regular interpolation sequence could therefore have a better cooperative cooling performance than the entire CM sequence. While the dependence of $\mathcal{C}$ for arbitrary hybrid sequence on its proportion of CM strategies might not be monotonic. We are then motivated to find an optimized sequence by virtue of the PPO algorithm. A typical RL-optimized sequence of cooling strategies labeled by $S_{\rm opt}$ is described in Fig.~\ref{CoolingPerformance}(h). With four orders reduction in the average population (close to the cooling efficiency provided by $S_c$), an almost unit ground-state fidelity $F>0.9999$, and a moderate success probability $P_g\approx30\%$ (much larger than that by $S_c$), the optimized sequence achieves an overwhelming cooperative cooling performance $\mathcal{C}(S_{\rm opt})=2.73$ according to Eq.~(\ref{cooperative}) over all the other measurement sequences. Therefore, we have achieved a compromise of cooling rate and success probability through the reinforcement leaning method with a much less overhead than the brute-force searching. The RL-optimized sequence is not unique, yet the current results of $\bar{n}$, $F$, $P_g$, and $\mathcal{C}$ in Fig.~\ref{CoolingPerformance} are almost invariant as long as there is one CM in the first several rounds.

\begin{figure}[htbp]
\centering
\includegraphics[width=0.95\linewidth]{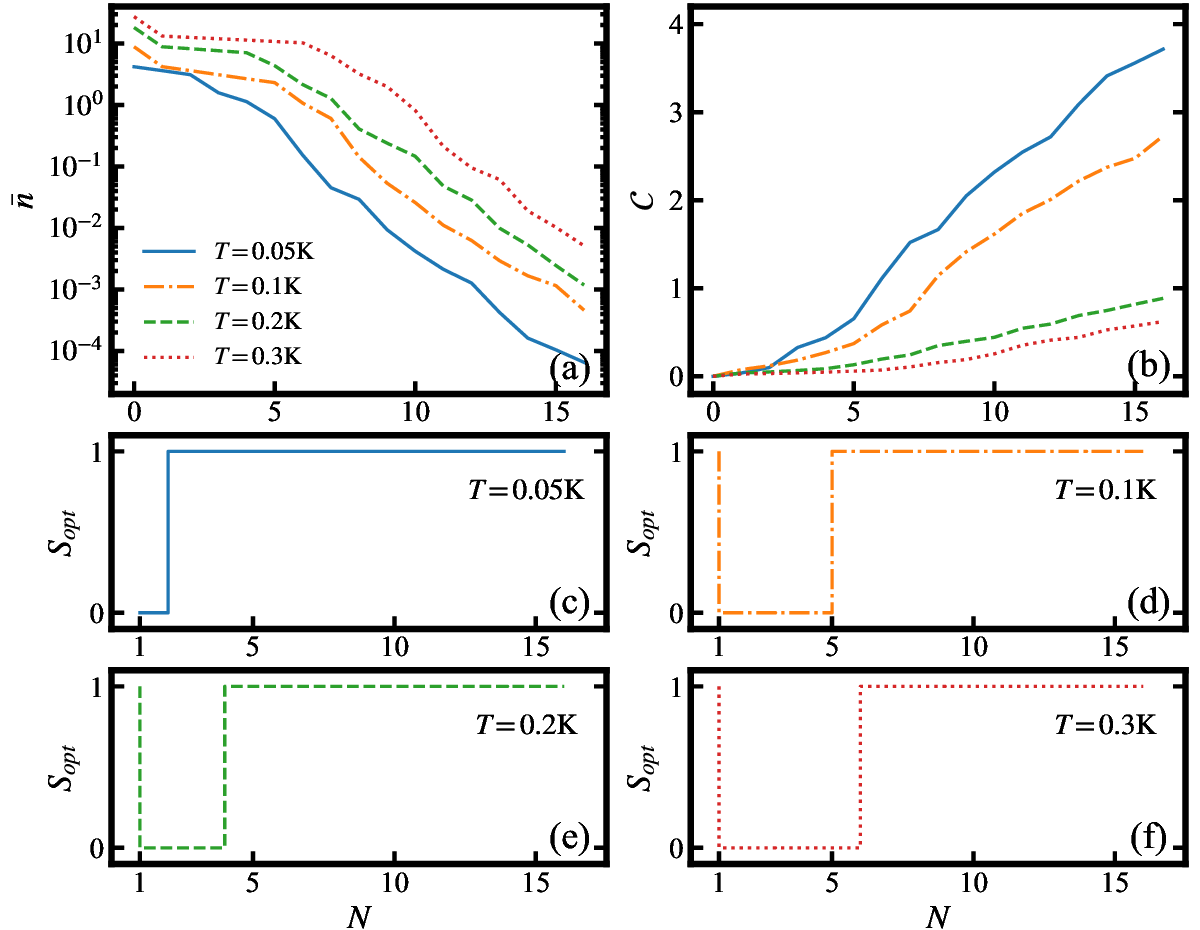}
\caption{(a) Average populations and (b) Cooperative cooling performance under the RL-optimized cooling algorithm with various initial temperatures. (c), (d), (e), and (f) describe the optimized sequences of UM and CM with $T=0.05$ K, $T=0.1$ K, $T=0.2$ K, and $T=0.3$ K, respectively. The other parameters are the same as those in Fig.~\ref{CoolingPerformance}. }\label{DifferentTemperature}
\end{figure}

The RL-optimized algorithm applies to a wide range of initial temperature for the resonator. Starting from various $\bar{n}_{\rm th}$ determined by the temperature, the average populations could be reduced by three to five orders in magnitude under the optimized measurement sequences, as demonstrated in Fig.~\ref{DifferentTemperature}(a). It is found that under a higher temperature, it is harder to suppress the transitions between the ground state and the excited states of the detector. Then both the relative magnitude in the population reduction [see Fig.~\ref{DifferentTemperature}(a)] and the cooperative cooling performance [see Fig.~\ref{DifferentTemperature}(b)] manifest a monotonically decreasing behavior as temperature increases.

Similar to Fig.~\ref{CoolingPerformance}(h), here we present in Figs.~\ref{DifferentTemperature}(c), (d), (e), and (f) the optimized sequences fully determined by the PPO algorithm, which still outperform any regular interpolated sequence in the cooling quantifier $\mathcal{C}$. Comparing these four sub-figures corresponding to various temperatures, it is interesting to find that a larger portion of the unconditional measurements is required along the optimized sequence for a higher temperature. It is consistent with the fact that under CM the success probability $P_g$ to find a detector in its ground state decreases exponentially with increasing temperature of the target resonator. Then more UMs are used to save a rapidly declining $P_g$ for obtaining a larger $\mathcal{C}$. In addition, for $T>0.05$ K, RL-optimized sequence always starts from a conditional measurement, which is important to have a significant cooling rate for $\bar{n}$ during the first several rounds of the whole sequence.

The profiles shown in Fig.~\ref{CoolingPerformance}(h) and Figs.~\ref{DifferentTemperature}(d), (e), and (f) manifest a common pattern for all the RL-optimized sequences. It is found in the previous several rounds that a conditional or projective measurement should be performed on the detector, when the resonator is normally in a comparatively high-temperature state, and several unconditional measurements ensued before further cooling. This pattern is consistent with the variations of both energy and entropy in nonunitary controls~\cite{MeasurementStabilizing}. The energy variation induced by a projective measurement is $k_BTH(\rho)$ on average, where $H(\rho)$ is the Shannon entropy of the whole system after a free evolution. Then in the end of the first round, a projective measurement is desired to cut down as much energy as it could, which is followed by several rounds of unconditional measurements to save the success probability. Thus in general we anticipate to see more UMs than CMs in the first several rounds and more CMs than UMs in the remaining rounds.

\section{Discussion and conclusion}\label{Conclusion}

\begin{figure}[htbp]
\centering
\includegraphics[width=0.95\linewidth]{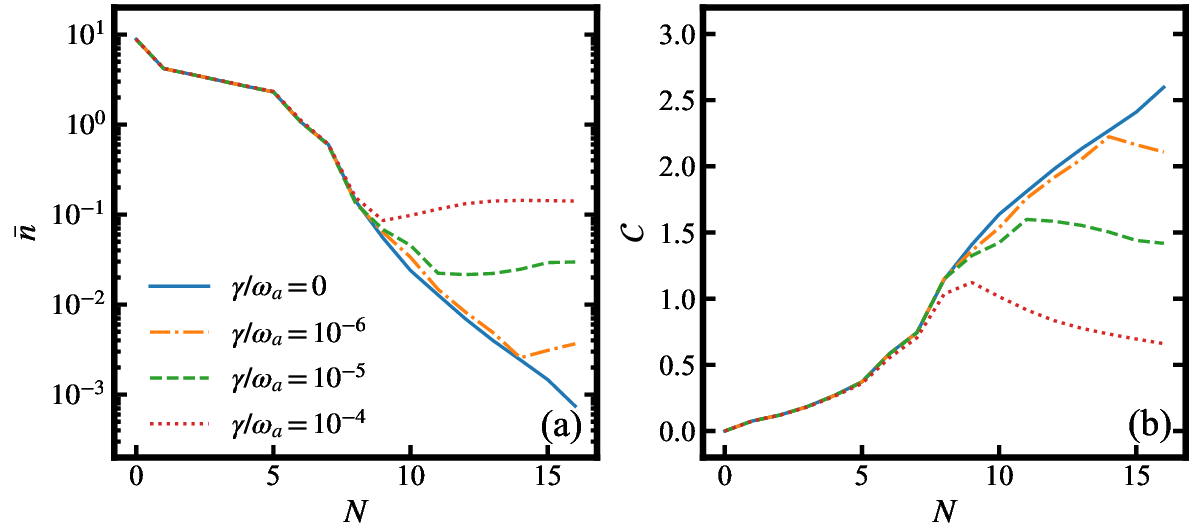}
\caption{(a) Average population and (b) Cooperative cooling performance of the resonator coupled to a thermal environment under the optimized cooling strategy with various dissipative rates. The dissipation-free results are those labeled by $S_{\rm opt}$ in Figs.~\ref{CoolingPerformance}(a) and (d).}\label{EnvironmentEffect}
\end{figure}

Preceding analysis over the cooling performance neglects the environment-induced dissipation. We now consider the cooling process in an open-quantum-system scenario, in which the free evolution between neighboring measurements is influenced by a finite-temperature environment. The dynamics is then described by the master equation
\begin{equation}\label{ME}
\begin{aligned}
\dot\rho(t)=&-i[H, \rho(t)]\\ &+\gamma(\bar{n}_{\rm th}+1)\mathcal{D}[a]\rho(t)+\gamma\bar{n}_{\rm th}\mathcal{D}[a^\dagger]\rho(t),
\end{aligned}
\end{equation}
where $\mathcal{D}[A]$ represents the Lindblad superoperator
\begin{equation}\label{Lindblad}
\mathcal{D}[A]\rho(t)\equiv A\rho(t)A^\dagger-\frac{1}{2}\left\{A^\dagger A, \rho(t)\right\}.
\end{equation}
In Fig.~\ref{EnvironmentEffect}(a) and (b), we present the average population $\bar{n}$ and the cooperative cooling performance $\mathcal{C}$ respectively with various dissipation rates. To compare the cooling performances in the presence of thermal decoherence to the dissipation-free situation, we apply the RL-optimized sequence provided in Fig.~\ref{CoolingPerformance}(h). It is found that a larger dissipation rate gives rise to a weaker cooling performance in terms of both $\bar{n}$ and $\mathcal{C}$, exhibiting the struggle between cooling effects by measurement and the accumulated heating effects by environment. Nevertheless, for typical mechanical resonators in gigahertz with $\gamma/\omega_a\sim10^{-5}$~\cite{MechanicalResonator1,MechanicalResonator2}, our optimized cooling protocol is still capable to reduce $\bar{n}$ by three orders in magnitude with about $N=10$ measurements [see the green dashed line in Fig.~\ref{EnvironmentEffect}(a)]. In the mean time, the asymptotic value of $\mathcal{C}$ still overwhelms the CM strategy labeled by $S_c$ in Fig.~\ref{CoolingPerformance}(d).

Even in the absence of thermal decoherence, $\bar{n}$ does not keep decreasing. Fundamentally, it is under the constraint of the third law of thermodynamics, that the absolute zero cannot be attained within a finite number of operations. Actually, either $\tau_{\rm opt}^c$ or $\tau_{\rm opt}^u$ approaches infinity as $\bar{n}\rightarrow0$, which indicates that the whole cooling process has to be truncated by a maximum timescale.

We emphasize again that the preceding hybrid cooling sequences based on the conditional and unconditional measurements are optimized in both global and local perspectives. Globally, we use the reinforcement learning to find the optimized order for UM and CM. The local optimization depends on the selected measurement interval to obtain a minimum average-population $\bar{n}$ under one measurement. For UM in Eq.~(\ref{Optimaltau}), $\tau_{\rm opt}^u(t)$ is not necessarily obtained by an instant feedback mechanism during a realistic practice. The measurement sequence $\{\tau_1(t_1), \tau_2(t_2), \cdots, \tau_N(t_N)\}$ can be actually obtained prior to the cooling measurements. $\tau_1(t_1)$ depends on the initial population-distribution $p_n$, and $\tau_k(t_k)$, $k\geq2$, can be calculated on the effective temperature that is uniquely determined by the dynamics of $p_n(t)$ through Eq.~(\ref{Omegad}). In other words, we can avoid the feedback error and imprecision induced by detecting the resonator states during the experiment.

In summary, we present an optimized cooling architecture on a sequential arrangement of both conditional and unconditional measurements. We analyse and compare the advantages and disadvantages of both CM and UM on cooling rate and success probability. We obtain analytically for the first time an analytical expression for the optimized unconditional measurement-interval $\tau_{\rm opt}^u=\pi/(\Omega_d+\Omega_{d+1})$ in parallel to that for conditional measurement~\cite{TwoModeCooling}. Here the dominant Rabi frequency $\Omega_d$ depends on the dominant distribution of resonator in its Fock state with $n_d=k_BT/(\hbar\omega_a)$ and the coupling strength between target and detector. The combination of the advantages of both measurement strategies gives rise to an optimized hybrid cooling algorithm assisted by the reinforcement learning. It is justified by the cooperative cooling performance as we defined to quantify the comprehensive cooling efficiency for arbitrary cooling-by-measurement strategy. Our work therefore pushes the cooling-by-measurement to an unattained degree in regard of efficiency and feasibility. It offers an appealing interdisciplinary application of quantum control and artificial intelligence.

\section*{Acknowledgments}

We acknowledge financial support from the National Science Foundation of China (Grants No. 11974311 and No. U1801661).

\appendix

\section{Proximal Policy Optimization}\label{PPOSec}

This appendix provides more details in proximal policy optimization, a typical reinforcement learning algorithm that we use to optimize the measurement sequence for cooling. PPO algorithm follows an ``actor-critic'' frame, in which actors receive the current state as an input and then outputs an action according to an updatable policy, and a critic evaluates this action to determine whether the action should be encouraged or not. In the following, we do not discriminate ``actor" and ``policy'' for simplicity.

\begin{figure}[htbp]
\centering
\includegraphics[width=0.95\linewidth]{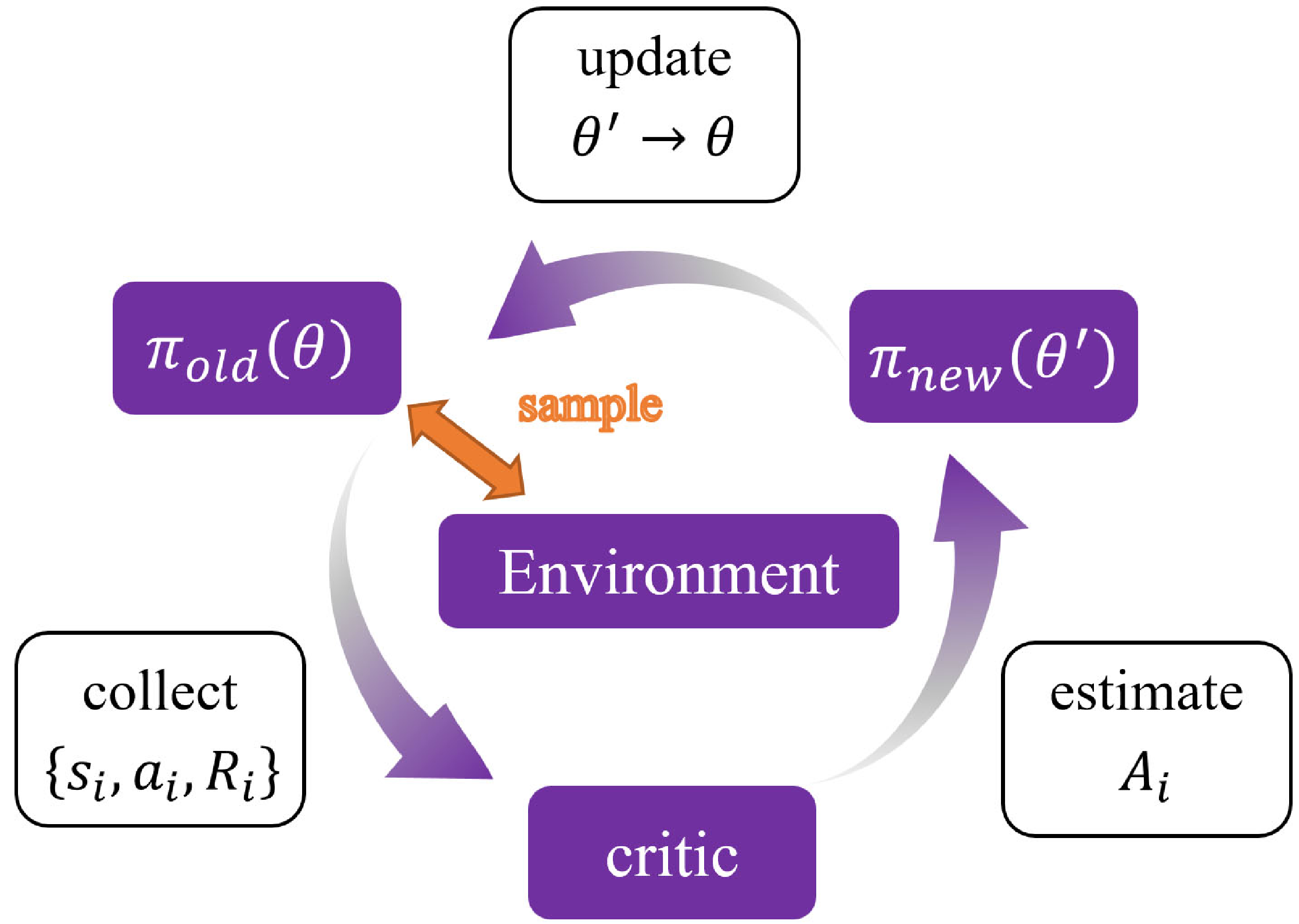}
\caption{Diagram of proximal policy optimization algorithm.}\label{PPO}
\end{figure}

As shown in Fig.~\ref{PPO}, PPO algorithm has two actors (policies) $\pi_{\rm old}(\{\theta\})$ and $\pi_{\rm new}(\{\theta'\})$ and one critic. Any of them is of an agent constructed by the neural networks (see Fig.~\ref{Model}) feathered with a set of parameter $\{\theta\}$. The two policies have the same structures in PPO. The old policy collects the sampling data through interaction with the environment; and the new one would use these data stored in a buffer to update $\{\theta\}$ to be $\{\theta'\}$. At first, the environment would initialize and deliver the state $s_1$ of the target system to the old policy $\pi_{\rm old}(\{\theta\})$; then the old policy generates an action $a_1$ according to $s_1$ and $\{\theta\}$. In environment, the action $a_1$ is taken and the system state becomes $s_2$. The environment also provides a reward $R_1$ indicating how good the action is. The reward is generated by a task-specified reward function. At this stage, an interaction between the policy and the environment is completed and one set of ``trajectory'' or return $\{s_1, a_1, R_1\}$ is collected. $N$ trajectories are collected in one episode, where $N$ amounts to the number of actions required to complete the task. The critic takes both actions and states as input and outputs an advantage $A_i$ representing the contribution of the current action $a_i$ on the current state $s_i$. After collecting a sufficient amount of data, the critic would estimate the actions' contribution as precise as possible. In the mean time, according to the advantages to maximize a clipped surrogate objective function $L^{\rm CLIP}(\{\theta\})$~\cite{PPO}, the new policy would transfer its parameters $\{\theta'\}$ to the old one.

In our application for optimizing the cooling sequence, the allowed inputs of the system states are defined as the populations in the Fock states, i.e., the diagonal elements of the target resonator $\rho_a$
\begin{equation}
s_i=\{p_0(t), p_1(t), p_2(t), \cdots, p_{n_c}(t)\},
\end{equation}
where $n_c$ indicates the cutoff Fock-state for the resonator. The actions taken by the environment are selected from the set
\begin{equation}
a_i\in\{0, 1\},
\end{equation}
where $0$ and $1$ represent unconditional and conditional measurements, respectively. Two policies are used to decide which type of measurement to be performed due to the current state of the resonator. Environment represents the quantum devices performing measurements, obtaining the updated states, and returning the rewards. When an action is selected and sent to the environment, the optimized measurement interval is calculated according to the measurement type. After unitary evolution lasting $\tau_{\rm opt}\in\{\tau_{\rm opt}^c, \tau_{\rm opt}^u\}$, measurement is performed on the detector. Then the average population $\bar{n}$, the ground-state fidelity $F$, and the success probability $P_g$ are obtained to calculate the cooperative cooling performance $\mathcal{C}$ given by Eq.~(\ref{cooperative}). The reward function is set as a certain multiple of $\mathcal{C}$, $R_i(s_i, a_i)=100\times\mathcal{C}(s_i,a_i)$. After measurement, the environment then returns the resonator state and the reward to the policies. When the training is completed, a policy $\pi(\{\theta_{\rm opt}\})$ with a set of optimized parameters is achieved. The neutral network equipped with $\{\theta_{\rm opt}\}$ could then be used to generate the optimized actions to cool down the current state.

\section{Generation of optimized sequence}\label{OptSequence}

\begin{algorithm}[H]
\SetAlgoNoLine
\SetKwFor{For}{for}{do}{end}
\SetKwIF{If}{ElseIf}{Else}{if}{then}{else if}{else}{\quad end}
\KwOut{$S_{\rm opt}=\{\mathcal{M}_1, \mathcal{M}_2, \cdots, \mathcal{M}_N\}$ and $\mathcal{T}=\{\tau_{\rm opt}(t_1), \tau_{\rm opt}(t_2), \cdots, \tau_{\rm opt}(t_N)\}$}
\KwIn{Temperature $T$}
Initialize the thermal state $\rho_a=\sum_np_n|n\rangle\langle n|$ with $T$
Use PPO to train an optimized policy $\pi(\{\theta_{\rm opt}\})$  \\
\For{$i=1, 2, \cdots, N$}{
\quad Run the policy $\pi(\{\theta_{\rm opt}\})$ on $\rho_a$ to generate $\mathcal{M}_i$ \\
\quad Attain $T_{\rm eff}=\hbar\omega_a/[k_B\ln{(1+1/\bar{n}})]$ on $\bar{n}(\rho_a)$ \\
\quad \If{$\mathcal{M}_i=0$}{
\qquad Calculate $\tau_{\rm opt}(t_i)=\pi/(\Omega_{d}+\Omega_{d+1})$ on $T_{\rm eff}$ \\
\qquad Get the cooling coefficients $|\alpha_n|^2$ and $|\beta_n|^2$ \\
\qquad UM: $\rho_a\gets\sum_n(|\alpha_n|^2p_n+|\beta_{n+1}|^2p_{n+1})|n\rangle\langle n|$ \\
}
\quad \ElseIf{$\mathcal{M}_i=1$}{
\qquad Calculate $\tau_{\rm opt}(t_i)=1/\Omega_{\rm th}$ on $T_{\rm eff}$ \\
\qquad Get the cooling coefficients $|\alpha_n|^2$  \\
\qquad CM: $\rho_a\gets\sum_n|\alpha_n|^2p_n|n\rangle\langle n|/(\sum_n|\alpha_n|^2p_n)$}
}\caption{RL-optimized cooling procedure} \label{Sequence}
\end{algorithm}

Both the order of measurements and the sequence of measurement-intervals could be regarded as output of our RL-optimized cooling algorithm as shown in Algorithm~\ref{Sequence}. The input information is the initial temperature $T$, fully determining the thermal state of the resonator. When the reinforcement learning process was completed by PPO algorithm (see Appendix~\ref{PPOSec}), the parameters $\{\theta\}$ of the neural network (policy $\pi$) have been trained to be capable to select one of the two measurement strategies for the current state, which maximizes the cooperative cooling performance. And then the cooling procedure is formally launched. We run the policy $\pi(\{\theta_{\rm opt}\})$ on $\rho_a(0)=\rho_a^{\rm th}$, which generates the first measurement strategy $\mathcal{M}_1$, $\mathcal{M}_1\in\{0, 1\}$. Here $0$ and $1$ indicate UM and CM, respectively. If $\mathcal{M}_1=0$, then $\tau_{\rm opt}(t_1)=\tau_{\rm opt}^u$ in Eq.~(\ref{Optimaltau}) that could be obtained by the effective temperature $T_{\rm eff}$ of the resonator (initially $T_{\rm eff}=T$, and it is updated by the current state of the last round). Subsequently, the cooling coefficients $|\alpha_n|^2$ and $|\beta_n|^2$ are calculated and the resonator state is modified according to Eq.~(\ref{umrhoa}). Otherwise if $\mathcal{M}_1=1$, a conditional measurement will be implemented after an interval $\tau_{\rm opt}(t_1)=\tau_{\rm opt}^c=1/\Omega_{\rm th}(t)$ and the resonator state is modified according to Eq.~(\ref{rhoace}). In the end of this round, one can calculate $T_{\rm eff}$ by the current $p_n(t)$ and then go to the next round. After $N$ iterations, the optimized measurement sequence characterized by $S_{\rm opt}=\{\mathcal{M}_1, \mathcal{M}_2, \cdots, \mathcal{M}_N\}$ and $\mathcal{T}=\{\tau_{\rm opt}(t_1), \tau_{\rm opt}(t_2), \cdots, \tau_{\rm opt}(t_N)\}$ appear as respectively described in Fig.~\ref{CoolingPerformance}(h) and Figs.~\ref{DifferentTemperature}(c), (d), (e), (f). In practical implementations, the measurements by $S_{\rm opt}$ and $\mathcal{T}$ can be acted on the detector without knowledge of the target-resonator state.

\bibliographystyle{apsrevlong}
\bibliography{ref}

\end{document}